\documentstyle[referee]{l-aa}
\newcommand{\gapr}{\raisebox{-.6ex}{\mbox{
$\stackrel{>}{\mbox{\scriptsize$\sim$}}\:$}}}
\newcommand{\lapr}{\raisebox{-.6ex}{\mbox{
$\stackrel{<}{\mbox{\scriptsize$\sim$}}\:$}}}
\def\Tef{T_{\rm eff}}
\def\ns{1E~1207.4--5209}
\def\nh{n_{H,21}}
\def\ros{{\sl ROSAT}}
\def\asca{{\sl ASCA}}
\begin{document}
\thesaurus{02.          
              (08.14.1;   
               08.19.5;   
               02.18.6;   
               13.25.5)   
}
\title{The Neutron Star in the Supernova Remnant PKS 1209--52}

\author{V.~E. Zavlin\inst{1,2}, G.~G. Pavlov\inst{2} and J. Tr\"umper\inst{1}}
\institute{Max-Planck-Institut f\"ur Extraterrestrische Physik, D-85740
Garching, Germany
\and
The Pennsylvania State University, 525 Davey Lab,
University Park, PA 16802, USA}
\date\today 
\maketitle
\markboth{Zavlin et al.: Neutron Star in PKS 1209--52}{}
\begin{abstract}
We re-analyzed soft X-ray data collected with the \ros\ and \asca\
observatories on a candidate neutron star (NS) near the center of the
supernova remnant PKS 1209--52. We fitted the observed spectra
with NS atmosphere models.
The hydrogen atmosphere fits yield more
realistic parameters of the NS and
the intervening hydrogen column than
the traditional blackbody fit. In particular,
for a NS of mass $1.4~M_\odot$ and radius 10~km,
we obtained the NS surface
temperature $T_{\rm eff}=(1.4-1.9)\times
10^6$~K
and distance $d=1.6-3.3$~kpc
versus $T=(4.2-4.6)\times 10^6$~K
and (implausible) $d=11-13$~kpc
for the blackbody fit, at a 90\% confidence level. Our fits
suggest that the surface magnetic field
is either very weak, $B\lapr 10^{10}$~G, or it exceeds $\simeq 2\times
10^{12}$~G.
The hydrogen column density inferred from the atmosphere fits, $n_H=
(0.7-2.2)\times 10^{21}$~cm$^{-2}$,
agrees fairly well with independent estimates obtained from UV
observations of nearby stars, radio data, and X-ray spectrum
of the shell of the supernova remnant, whereas the blackbody and
power-law fits give considerably lower and greater values,
$n_H=(0.2-0.4)\times 10^{21}$ and $(5.2-7.0)\times 10^{21}$~cm$^{-2}$,
respectively.
The inferred NS surface temperature is consistent with standard
NS cooling models.
\keywords{
stars: neutron: individual (\ns) --- supernova remnants: individual
(PKS 1209--52) --- X-rays: stars}
\end{abstract}
\section{Introduction}
In addition to about 15 radio pulsars associated with supernova
remnants (Gaensler \& Johnston 1995), 
several {\it radio silent} isolated neutron star
(NS) candidates within SNRs have been 
observed with the X-ray observatories {\sl HEAO} A, {\sl Einstein}, 
{\sl EXOSAT}, \ros, and \asca\
(see Caraveo, Bignami \& Tr\"umper 1996 for a review). 
These objects have not been detected outside the X-ray range,
and their X-spectra resemble blackbody (BB) spectra with temperatures
of a few million kelvins.
If they are indeed thermally emitting NSs,
the analysis of their radiation provides an opportunity to study
thermal evolution of NSs of ages $\sim 10^3-10^5$~yr, which is 
important
for elucidating the properties of the superdence matter in 
NS interiors.

One of the most convincing examples of such objects is the point-like
source \ns\ within the barrel-shaped radio, X-ray, and 
optical SNR PKS 1209--52 (also known as G269.5+10.0). From
the analysis of radio and optical observations of this SNR,
Roger et al.~(1988) estimated its age $\sim 7000$~yr,
with an uncertainty of a factor of 3, and concluded that the
remnant is in an adiabatic expansion phase. 
The distance to PKS 1209--52  is not well determined --- estimates
in the range $1.1-3.9$~kpc were suggested (Milne 1979; Mills 1983).
Estimates of the interstellar hydrogen column density from the
radio and optical data yield $\nh\equiv n_H/(10^{21}~{\rm cm}^{-2})
\sim 1.0-1.8$
(see Roger et al.~1983 and Kellet et al.~1987 for references),
consistent with a distance $d\sim 1-2$~kpc.  

After the first X-ray detection of
PKS 1209--52 with {\sl HEAO}~A-1 (Tuohy
et al.~1979), the point source \ns\ was discovered 
with the {\sl Einstein} observatory (Helfand \& Becker 1984),
$6'$ off-center the $81'$ diameter SNR. Matsui,
Long \& Tuohy (1988) concluded that its spectrum can be interpreted
as a BB spectrum with an apparent temperature
$T=1.4\times 10^6$~K (assuming $n_{H,21}=3.2$,
as obtained from the {\sl HEAO}~A-1 observations).
The {\sl Einstein} High Resolution Imager (HRI) observations showed 
a lack of diffuse X-ray emission around \ns~ (later confirmed
by the \ros\ HRI observations),
which greatly simplifies the analysis of the point source radiation. From
the analysis of {\sl EXOSAT} observations,
Kellett et al.~(1987) estimated 
the BB temperature of the central object $T= 1.8\times 10^6$~K
(at $\nh = 1.1$), which implies an emitting area of a radius
$(3-4)~d_2$~km, where $d_2=d/(2~{\rm kpc})$.
Applying the Raymond \& Smith (1977) line-emission model,
Kellet et al.~estimated also an average SNR temperature,
$\sim 1.7\times 10^{6}$~K, and the hydrogen column towards the SNR,
$\nh\simeq 1.4$, consistent
with the values obtained from radio and optical data. 

Observations of PKS 1209--52 and its central source with 
\ros\ and \asca\  
have further supported the NS hypothesis for \ns.
Mereghetti, Bignami \& Caraveo (1996; hereafter MBC96) 
showed that the \ros\ 
data on \ns~ can be interpreted as blackbody emission of $T\sim 
3\times 10^6$~K from an area with radius $R\sim 1.5~d_2$~km. 
The authors noticed that this
temperature is too high to be explained in the framework
of a cooling NS with age of PKS 1209--52 ($\sim 10^4$ yr). 
The hydrogen column inferred from the BB fit,
$\nh\sim 0.4$,
is 3--4 times lower than the estimate given by Kellett et al.~(1987). From
observations at 4.8 GHz, MBC96 found an upper limit
of $\sim 0.1$~mJy on the radio flux from \ns.
They also set a deep limit of $V > 25$ for an optical counterpart 
in the {\sl Einstein} HRI error circle,  
that supports the hypothesis that
\ns~ is indeed an isolated NS. 

Vasisht et al.~(1997; hereafter Val97) have recently analysed \asca\
observations of \ns.  They found that
each of the three spectra obtained with the \asca\ detectors
can be fitted by a BB spectrum consistent with that obtained
from the analysis of the \ros\ data.
The hydrogen column is poorly restricted
in the \asca\ spectral fits due to low detector
sensitivities at photon energies below 0.5~keV. Using a 
fit with a Raymond-Smith model at fixed cosmic abundances
to the \ros\ data on the SNR shell,
Val97 estimated the remnant temperature $\sim (1.9-2.3)\times 10^6$~K
and the foreground column density $\nh\sim 0.6 - 0.9$.
They attributed the difference in $n_H$ between
the NS and SNR fits to either 
separate lines of sight or the large column density that
the shell X-rays can encounter in the SNR postshocked gas.

Based on the results of the blackbody analysis, both MBC96
and Val97 concluded that \ns\ is an isolated NS with
hot spots on its surface, aligned with the magnetic poles. 
Val97 suggested that the spots are heated either by
dissipative heating in the NS interior or by the bombardment
of polar caps by relativistic particles from the NS magnetosphere
if \ns\ is an active pulsar.
However, the former hypothesis can hardly explain the small
sizes of the hot spots, even 
with  allowance for large anisotropy
of thermal conductivity of the magnetized NS crust.
The latter heating
mechanism is also in doubt because of
the absence of radio and $\gamma$-ray emission from \ns. 
Moreover, both the \ros\ and \asca\ data did not reveal pulsations,
which may not be consistent with the presence of the hot spots
unless the magnetic and rotation axes are coaligned.

Although it looks very plausible that
\ns\ is a thermally emitting 
isolated NS, the 
BB interpretation adopted by previous authors leaves several 
controvertible points. On the other hand,
thermal radiation emitted by 
NS atmospheres may significantly differ from the BB radiation.
Therefore, to resolve the inconsistencies following from the
simplified BB interpretation, it is natural to employ 
more realistic models of NS radiation.
Here we present a {\it combined} analysis of the \ros\ and \asca\ data
based on {\it NS atmosphere models} (Pavlov et al.~1995,
and references therein). These models have been applied 
successfully to the interpretation of the soft X-ray radiation
from, e.g., the Vela pulsar (Page, Shibanov \& Zavlin 1996) and
the brightest millisecond pulsar J0437--4715 (Zavlin \& Pavlov 1997;
Pavlov \& Zavlin 1997). These examples show that fitting the
soft X-ray pulsar spectra with hydrogen or helium
atmosphere models always
results in lower effective temperatures and greater emitting
areas (or smaller distances) than those obtained from the BB fits.

We show in \S 2 that, indeed, the model atmosphere fits of the
X-ray radiation from \ns\ 
yield an NS surface temperature
compatible with NS cooling models, and they do not require hot spots
on the NS surface.
Moreover, the hydrogen column density inferred from this interpretation
is in excellent agreement with that obtained from our fits of the
SNR X-ray radiation as well as with independent estimates of $n_H$ for
stars in the vicinity of \ns\ (\S 3). 
These results warrant application of the same approach to other
similar objects and enable one to obtain reliable estimates of
surface temperatures of NSs of different ages.

\section{Observations and results}
We used archival \ros\ and \asca\ 
data on \ns~ (see MBC96 and Val97 for a detailed description).
Four \ros\ Position Sensitive Proportional Counter
(PSPC) pointings (obtained in 1993 July, with exposure
times $\approx 5-6$~ks each --- see Table 2 in MBC96)
were centered at different positions
(off-axis angles between $16'$ and $33'$). 
We checked that the spectra of \ns~ extracted from 
each of the four data sets and properly
corrected for the PSPC response 
are consistent with each other.
For the combined analysis with the \asca\ data
we chose the \ros\ PSPC observation with the minimum off-axis angle $16'$
(exposure time $4.97$~ks).
We extracted a raw spectrum from a region with
a radius $r\simeq 75''$ centered on the source.
The (off-axis) source count rate is $0.145\pm 0.005$~s$^{-1}$.
We binned this spectrum into 24 spectral bins in the
0.1--2.4~keV energy range for further analysis.
The \ros\ HRI counts for each of the three on-axis pointings
(obtained in 1992 August and 1994 July; total exposure time
$\approx 14.5$~ks) were selected from
the circles of $r\simeq 22''$. The corresponding source count rate is 
$0.064\pm 0.003$~s$^{-1}$, in agreement with MBC96.

The \asca\ data were obtained in 1994 July, with two
Solid State Imaging Spectrometers, SIS0 and SIS1,
in the bright mode, and two Gas Scintillation
Imaging Spectrometers, GIS2 and GIS3, in the pulse-height mode.
The SIS1 image of \ns~ is too close to the edge of the CCD chip
for a reliable spectral analysis.
The SIS0 source counts collected in 
$\approx 21$~ks (with high and
medium bit rates) were extracted from the circle of $r\simeq 4'$,
the total source count rate is $0.0556\pm 0.0017$~s$^{-1}$. 
The SIS0 spectrum was binned into 38 spectral bins 
in the $0.5-10.0$~keV range.
The GIS2 and GIS3 counts were extracted from circles of
$r\simeq 9'$ (exposure 
$\approx  19$~ks for the high plus medium bit rates).
The source count rates are $0.0425\pm 0.0018$ (GIS2) and
$0.0533\pm 0.0025$~s$^{-1}$ (GIS3).
The spectra were binned into 40 and 45 spectral bins in
the $0.3-5.0$~keV range for GIS2 and GIS3, respectively.
The total source count rates 
for all the three \asca\ instruments are somewhat
lower  than those reported by Val97,
perhaps because of different screening criteria we applied. 

As a first step of our data analysis, we fitted the count rate spectra
{\it simultaneously
for the four instruments}, \ros\ PSPC and \asca\ SIS0, GIS2,
and GIS3, with the traditional power-law and BB models.
Figure~1 shows the results of these fittings.
The power-law fit yields
a photon index $\gamma = 5.2(+0.4,-0.3)$ and 
a hydrogen column density, $\nh = 6.0(+1.0,-0.8)$
(minimum $\chi_\nu^2=1.47$, $\nu=144$;
the uncertainties hereafter are
at a 90\% confidence level). These values of $\gamma$ and $n_H$
are better constrained than those obtained by MBC96 from the 
PSPC data alone, $\gamma=3.9(+1.5,-1.4)$ and $\nh = 3.0(+2.5,-1.8)$
(minimum $\chi_\nu^2=1.61$, $\nu=10$). The photon index is unusually large
in comparison with typical values, $\gamma \sim 1.5-2.5$, observed
from, e.g., X-ray emitting radio pulsars 
(Becker \& Tr\"umper 1997), 
and the hydrogen column density too much exceeds the values obtained
by independent measurements (see below). These discrepancies,
together with the too high minimum $\chi_\nu^2$, imply that the
power-law interpretation is inadequate.

The BB temperature (as measured by a distant observer), 
$T^\infty=3.12 (+0.15, -0.11) \times 10^6$~K, 
the apparent radius of the emitting area, $R_a  
=1.21 (+0.10, -0.13)~d_2$~km,
and the bolometric luminosity, 
$L_{\rm bol}^\infty = 0.93 (+0.19,-0.11)\times 10^{33}~d_2^2$~erg~s$^{-1}$,
obtained from the combined fitting (Fig.~1)
are very close to the values from
the separate fits of the  \ros\ and \asca\ data by MBC96 and Val97. 
The hydrogen column density in the combined fitting,
$\nh =0.28 (+0.15, -0.12)$,
is compatible with (albeit slightly lower than) the value of 
$\nh = 0.3-0.5$ inferred from the separate \ros\ PSPC spectrum. 
Although the quality of the BB fit is better than of the power law
fit, the $n_H$ values look surprisingly low in comparison with
$(1.0-1.8)\times 10^{21}$~cm$^{-2}$ obtained from independent estimates
(see Introduction).

Since thermal NS spectra, as well as spectra of ordinary stars, 
cannot exactly coincide with the
BB spectra because of the effects of radiative transfer in the
emitting layers, it is natural to compare them with more realistic model
spectra of NS atmospheres (e.g., Pavlov et al.~1995). Strongest
deviations from the BB spectrum are expected if the NS surface is 
covered with a hydrogen atmosphere. Hydrogen may appear at the NS
surface as a result of, e.g., accretion of interstellar matter
or post-supernova accretion of a fraction of the ejected envelope.
Due to the huge surface gravity (typical gravitational acceleration  
$\sim 10^{14}-10^{15}$~cm~s$^{-2}$), heavier chemical elements sink down in
deeper layers and do not affect properties of emitted radiation,
whereas hydrogen remains at the surface.
The shape of the spectrum emitted by an atmosphere depends on
the strength $B$ of the NS surface magnetic field.
The lack of pulsations does not allow one to estimate $B$
using period and its derivative, as is usually done for pulsars.
The absence of statistically reliable features in the observed spectra,
which could be associated with an electron cyclotron line at
$E_{Be}=11.6 (B/10^{12}~{\rm G})$~keV,
prevents one to make a direct estimation of $B$.
Therefore, one has to try the atmosphere models with both
low ($B \lapr 10^{10}$~G) and high magnetic fields.
(In the former case, the field
does not affect properties of emergent radiation [Zavlin, Pavlov \&
Shibanov 1996], so that we can merely put $B=0$).
In the case of strong nonuniform field ($B\gapr 10^{12}$~G), 
the NS should have a nonuniform
temperature distribution along its surface
because of the high anisotropic thermal conductivity in the NS crust
(e.g., Shibanov \& Yakovlev 1996). However, since we have no
information about the field geometry, we assume that the magnetic
field has the same strength and is directed radially everywhere at
the surface, and the effective temperature is uniform.
This assumption reduces the number of fitting parameters and can
be considered as a reasonable first approximation for investigating
the atmosphere effects. 

Figure~2 shows results of 
fitting of the combined \ros\ and \asca\ data with 
hydrogen atmosphere models for three fixed values of the magnetic
field. In these fits we assume canonical values for the NS mass
and radius, $M=1.4 M_\odot$ and $R=10$~km, and consider 
the distance as a fitting parameter.  The models with $B=0$ (left panels)
and $10^{13}$~G (right panels) result in close values of the distance,
effective temperature at the NS surface and bolometric luminosity,
$d= 2.3 (+1.0,-0.7)$~kpc, $\Tef=1.63 (+0.27,-0.19)\times 10^6$~K,
$L_{\rm bol}=5.0(+4.3,-1.8)\times 10^{33}$~erg~s$^{-1}$,
and $d=2.3 (+1.0,-0.8)$~kpc, $\Tef =1.65 (+0.20, -0.17) \times 10^6$~K,
$L_{\rm bol}=5.3(+3.0,-2.0)\times 10^{33}$~erg~s$^{-1}$,
for $B=0$ and $10^{13}$~G, respectively. 
Since the nonmagnetic spectra are softer at lower energies, the 
hydrogen column density at $B=0$, $\nh =1.0 (+0.5, -0.6)$,
is lower than for the strong magnetic field, $\nh =1.5(+0.7, -0.6)$.
We checked that folding the models 
within the 90\% confidence region with the
\ros\ HRI response yields count rates compatible with those
observed.  When the field strength varies in the range 
$5\times 10^{12}\lapr B \lapr 5\times 10^{13}$~G,
the fitting parameters remain approximately the same 
because the model spectra are almost insensitive to the $B$ value
in the corresponding domain of energies and effective temperatures.
When $B$ exceeds $\sim 5\times 10^{13}$~G, the proton cyclotron
line centered at $E_{Bp}=0.063 (B/10^{13}~{\rm G})$~keV
(cf.~Bezchastnov et al.~1996) enters
the SIS energy range, which makes the fits statistically unacceptable.
(This line moves above $\sim 5$~keV, a maximum energy where the NS
flux is still above the background, at superstrong magnetic fields,
$B\gapr 8\times 10^{14}~G$, for which the models we used here
are not directly applicable.)
When $B$ falls below $\simeq 5\times 10^{12}$~G,
the low-energy wing of the electron cyclotron line
gets into  the \asca\ range, and the model spectra become softer
at $E\gapr 2$~keV.
As a result, the confidence contours in the $n_H$--$d$ and $n_H$--$\Tef$
planes move to greater $d$ and $\Tef$, and lower $n_H$, towards the BB
contours. An example is shown in the middle panels of Fig.~2
for $B=2\times 10^{12}$~G, for which the best-fit distance
is about 50\% larger than at $B=10^{13}$~G.
When the field is lower than  $\sim 5\times 10^{11}$~G, but
greater than $\sim 1\times 10^{10}$~G, the atmosphere
model fits become statistically unacceptable because the core of
the electron cyclotron line gets into the \asca/\ros\ range.

The atmosphere models depend not only on $B$, but also on
the NS mass and radius which determine the gravitational acceleration
(one of our model parameters) and the gravitational redshift
factor $g_r=(1-0.295 M_*/R_{10})^{1/2}$, where
$M_*=M/M_\odot$, $R_{10}=R/10~{\rm km}$. To illustrate this effect,
we present in Fig.~3 the best-fit parameters at
$B=10^{13}$~G in a wide range of $R$ and $M$ allowed by equations of state 
of the NS matter. 
Although the effective temperature at the NS surface, $\Tef$, varies
by about $\pm 20\%$ in the allowed $R$-$M$ domain, the
apparent effective temperature (as measured by a distant observer),
$T_{\rm eff}^\infty=g_r \Tef$, remains almost constant,
$(1.23-1.31)\times 10^6$~K, because the redshift 
is compensated by the change of the unredshifted NS spectrum 
(it softens in the Wien tail with increasing $M$ and decreasing $R$
at given $\Tef$). 
Owing to the approximate constancy of $\Tef^\infty$, the
apparent and ``true'' bolometric luminosities depend on $R$
and $M$ as $L_{\rm bol}^\infty \propto R_\infty^2=R^2/g_r^2$
and $L_{\rm bol}=g_r^{-2}L_{\rm bol}^\infty\propto R^2/g_r^4$.
The latter dependence explains the non-monotonous behaviour of $L_{bol}$
at higher $M$.
The best-fit $d$ and $n_H$ are almost independent of $M$ at higher $R$, when
$g_r$ is close to 1. At lower $R$ the 
variations are stronger, albeit
within the statistical uncertainties of these parameters (cf.~Fig.~2).
The distance inferred at assumed $R$ and $M$ can be approximately
described, in the allowed mass--radius domain, by a linear equation:
$d=0.5 M_* - 0.4 +(2.5 - 0.3 M_*) R_{10}$ (in kpc); it grows with $R$ faster
than for the BB interpretation. Notice that if $d$ is determined
more accurately in future observations of the SNR and its central
source, this equation would delimit a band in the $M$--$R$ plane
constraining equation of state of the NS matter. 

Since the hydrogen atmosphere fits yield $n_H$ greater 
than the BB fit by a
factor of 3--5, it is important to estimate this parameter independently.
For this purpose, we fitted the SNR emission with various models
for thermal plasma radiation.
We extracted the SNR spectrum from
a bright region of the \ros\ PSPC image of the remnant shell within
the $r\simeq 5'$ circle centered at 
$\alpha_{(2000)}=12^{\rm h}11^{\rm m}56\fs 0$,  $\delta_{(2000)}=
-52^\circ 36' 36''$. Three models available with
the XSPEC package, vraymond, vmeka, and vmekal
(Raymond--Smith, Mewe--Gronenschild--Kaastra,
and Mewe--Kaastra--Liedahl models with variable abundances)
give satisfactory fits
($\chi^2_\nu < 1.0$) with the column density to the SNR
in the range of $(1.1-1.9)\times 10^{21}$~cm$^{-2}$, consistent
with the hydrogen atmosphere fits of \ns, but certainly in excess
of the BB fits. The inferred plasma temperature
is $(1.9-2.1)\times 10^6$~K. These results were obtained with a moderate
excess of abundances of Al and Si
whose emission lines are prominent in the SIS spectra of the SNR shell.
Abundances of other elements are close to standard 
values. The SNR parameters inferred from our fittings
are consistent with those obtained by Kellett et al.~(1987)
from the {\sl EXOSAT} data. The lower hydrogen density, $\nh = 0.6-0.9$,
obtained by Val97, is likely associated with fixed cosmic 
abundances adopted, that resulted in
lower statistical quality of their fits ($\chi^2_\nu=1.8$).

\section{Discussion and conclusions}
We have shown that the X-ray spectra of the isolated NS \ns~ in the
center of the SNR PKS 1209--52 observed with the \ros\ PSPC and HRI
and \asca\ SIS0, GIS2 and GIS3 can be interpreted 
as {\it thermal radiation from the 
hydrogen-covered, uniformly heated surface of the NS}.
The proposed interpretation resolves all the
inconsistencies which follow from 
the BB interpretation of radiation of this object.

{\it First}, the range of distances, $d=1.5-3.2$~kpc, inferred for 
the standard NS mass $M=1.4 M_\odot$ and radius $R=10$~km, is within the
(conservative) limits, $1.1-3.9$~kpc, obtained from radio
observations of the SNR. 
This range shifts to lower distances with decreasing $R$ and $M$:
e.g., $d=1.2-2.6$~kpc for $M=1.0 M_\odot$, $R=8$~km.
Thus, the assumption about hot spots on the NS surface,
which is difficult to reconcile with the lack of pulsar
activity in \ns, is superfluous in our model.

{\it Second}, the effective temperature 
$\Tef=(1.4-1.9)\times 10^6$~K ($\Tef^\infty=(1.0-1.6)\times 10^6$~K),
obtained from the fittings with our hydrogen atmosphere models,
matches well to a number of the NS cooling models (see Fig.~4
and discussion below). On the contrary, it is difficult to reconcile
the results of the BB interpretation with models of the NS cooling.

{\it Third}, the inferred
 hydrogen column density towards the \ns, $\nh = 0.9-2.2$
and $0.5-1.5$ for high and low surface magnetic fields, respectively,
agrees fairly well with $\nh =1.1-1.9$
obtained from fitting the spectra of the SNR shell. These estimates
are consistent with the $n_H$ values obtained by Kellett et al.~(1987)
from the {\sl EXOSAT} observations of PKS 1209--52.
Moreover, measurements of $n_H$ in UV observations
of several stars in the vicinity of \ns\ ($l=296.5^\circ$, $b=9.9^\circ$)
yield the hydrogen columns in virtually the same range
(Fruscione et al.~1994): $\nh = 1.1 -1.5$ for HD112244 
($l=303.6^\circ$, $b=6.0^\circ$, $d=1.85$~kps), $\nh = 1.4$
for HD115842 ($l=307.1^\circ$, $b=6.8^\circ$, $d=1.87$~kps)
and HD111822 ($l=303.1^\circ$, $b=10.2^\circ$, $d=2.53$~kps).
On the contrary, the BB fit gives $\nh = 0.15-0.45$,
clearly incompatible with all the other independent estimates.

The fact that \ns\ shows no pulsar activity can be explained,
in addition to the trivial explanation of unfavorable orientation of
its magnetic and rotational axes, by slow rotation or/and low
magnetic field of the NS, so that it falls below the ``death line''
on the $P$--$\dot{P}$ diagram,
$\dot{P}/P^3 =2\times 10^{-17}$~s$^{-3}$. For
the simplest model of the magnetic
dipole radiator, the pulsar period increases with time $t$ as
$P=(P_0^2 + 2 t_0 t)^{1/2}$, where $t_0=(8\pi^2/3)(B^2 R^6/I c^3)=
0.97\times 10^{-13} B_{13}^2 R_{10}^6 I_{45}^{-1}$~s, 
$I=10^{45} I_{45}$~g~cm$^2$ is the moment of inertia of the NS
and $B_{13}=B/(10^{13}~{\rm G})$.
The condition that the NS is below the death line can be written
as $t_0/P^4<2\times 10^{-17}$~s$^{-3}$, or 
$P>8.4 B_{13}^{1/2} R_{10}^{3/2} I_{45}^{-1/4}$~s. Hence, if
$B$ is in the range of strong magnetic fields
allowed by our fitting, $0.2\lapr B_{13} \lapr 8$, we may expect
that the NS rotates slowly, $P\gapr 1-2$~s. If faster pulsations
are discovered in future X-ray observations of \ns, it would indicate
that $B$ is in the other range compatible with our fits:
$B\lapr 10^{10}$~G. It is worth noting that
if the NS was born as a pulsar at an appreciable distance
from the death line, only an enormous magnetic field,
$B\gapr 5\times 10^{15}$~G at standard NS parameters, could decelerate
its rotation so that the pulsar would ``die'' at the
relatively young age, $t\sim 10^4$~yr. Hence, discovery of slow
pulsations would mean that the NS was born slowly rotating unless
its magnetic field is superstrong.

Since we have estimated the effective temperature of the NS, it
is interesting to compare it with what is predicted by various
cooling models. Figure 4 shows examples of NS cooling curves
from Van Riper, Link \& Epstein (1995) for
three equations of state (EOS), 
stiff (Pandharipande--Smith [PS]), intermediate 
(Friedman--Pandharipande [FP]) and soft (Baym--Pethick--Sutherland [BPS]),
and two interior compositions
resulting in slow and fast cooling (solid and dashed lines,
respectively). Cooling curves without additional heating are
depicted by thick lines, whereas thin lines show cooling curves
for two models, 
proposed by Epstein \& Baym (EB) and Alpar, Cheng \& Pines (ACP),
for pinning of the superfluid vortices to the
crust lattice. They correspond to strong and weak 
frictional heating associated with the dissipation
of energy of differential rotation between the NS crust and superfluid
interior (only the EB model is available for the BPS EOS).
In the same picture we
plot boxes corresponding to the inferred atmosphere and
BB temperatures of \ns, assuming its age in the range $(0.3-2)\times
10^4$~yr, and similar boxes for the Vela pulsar
whose X-ray radiation
was investigated in terms of the hydrogen atmosphere models by
Page et al.~(1996). (We adopted $(1-4)\times 10^4$~yr for the
Vela age,  between the conventional characteristic age $P/2\dot{P}$  
and an upper limit estimated by Lyne et al.~1996).
We see that the BB temperature of \ns\ is well above
the values predicted by all these cooling models, whereas
the BB temperature of the Vela pulsar is compatible
only with the standard (slow) cooling model supplemented by strong heating
for the stiff EOS. The effective temperature
of \ns\ obtained with the atmosphere fits
is compatible, given the poorly known age, with the 
slow cooling models for all the three EOS, at moderate or no
heating. For the
stiff EOS, it is also compatible with the fast (quark) cooling
accompanied by strong frictional heating. However, the large NS radius
for this EOS would mean a distance $d\simeq 2.5 - 5.2$~kpc (cf.~Fig.~3),
uncomfortably large in comparison with conventional estimates. The
same cooling curve (fast cooling with the EB heating for the 
stiff EOS) is the only one
which goes through both the \ns\ and Vela pulsar boxes obtained
from the atmosphere fits. This, however,
does not necessarily mean that slow cooling models or other EOS or other
heating rates are excluded. First, the \ros\ PSPC spectrum
of the Vela pulsar, used in both BB and atmosphere fits,
was not observed directly because the pulsar was not resolved
from a surrounding 
mini-nebula of $2'$ diameter, which makes the results
of the spectral fits less certain. Second, one cannot exclude,
in principle, that the NS of the Vela pulsar differs from
\ns\ (e.g., the mass of the former may be greater, that could
lead to an ``exotic'' interior composition associated with
an enhanced neutrino luminosity and accelerated cooling).
Finally, there exist many more cooling models than shown in Fig.~4, and
some of them may satisfy both the \ns\ and Vela pulsar temperatures.
For instance,
a strong neutrino emission induced by the nucleon Cooper pair 
formation process
(see Page 1997, and references therein), which was neglected in most
of the previous NS cooling models, 
may result in great variety of cooling
curves, depending on the (unknown) parameters of the nucleon pairing
(Yakovlev 1997). 
For a more detailed comparison of the inferred temperatures with 
the cooling models, it would be very important to evaluate more
accurately the distance to PKS 1209--52 (which would constrain
the NS radius and hence the EOS) and the age of this object.

There exist a number of other radio silent isolated NS candidates
whose observational manifestations are similar to those of \ns.
The most convincing example is 1E/RXJ 0820--4247 in the SNR Puppis A,
whose BB temperature, $\approx 3\times 10^6$~K, and radius, 
$\approx 2$~km (Petre, Becker \& Winkler 1996),
virtually coincide with those of \ns. Another example with similar
properties is RXJ 0002+6246 in the SNR G117.7+0.6 (Hailey \& Craig 1995), 
whose identification
is, however, less certain because of its faintness (likely, due to
a larger distance and greater $n_H$). We expect that applying the
NS atmosphere models to the analysis of such objects will allow us
to evaluate their radii and effective temperatures, and
to constrain their magnetic fields. 

One more radio silent NS candidate,
1E~1613--5055 in the center of the SNR RCW~103 (Tuohy \& Garmire 1980),
appears as a different kind of object ---  the BB fitting of its spectrum yields
a considerably higher temperature, $\sim 7\times 10^6$~K, at a few times smaller
radius of the emitting region (Gotthelf et al.~1997). However, this object
is deeply immersed in the remnant 
diffuse emission so that it is hard to separate
its spectrum from that of the SNR, with the limitied spatial resolution
of \asca. We expect that the forthcoming $AXAF$ and $XMM$ 
missions would resolve the point source and provide its spectrum
suitable for the detailed analysis.

\acknowledgements
The data were reduced and analyzed with MIDAS/EXSAS, FTOOLS/XSELECT
and XSPEC software packages.
We are grateful to Ken Van Riper for providing us with the
cooling curves in numerical form, to Dima Yakovlev for
clarifying various aspects of the NS cooling, and to Gordon
Garmire for the discussion of X-ray observations of isolated 
NSs in SNRs. The work was partially supported through
NASA grant NAG5-2807, INTAS grant 94-3834 and
DFG-RBRF grant 96-02-00177G.
VEZ thanks the Pennsylvania State University for hospitality
and acknowledges the Max-Planck fellowship.

\newpage
\centerline{\bf Figures}

{\bf Fig.~1}
68\%, 90\% and 99\% confidence contours for
the power-law and blackbody fittings to the
\ros\ PSPC and \asca\ SIS0, GIS2 and GIS3 spectra of \ns.
The lines in the right panel correspond to constant values
of radius of emitting area at a distance $d=2$~kpc.

{\bf Fig.~2}
68\%, 90\% and 99\% confidence contours for
the NS hydrogen atmosphere fittings for
$M=1.4 M_\odot$, $R=10$~km, and three values of
the surface magnetic field $B$ to the combined \ros\
and \asca\ data.

{\bf Fig.~3}
Dependences of the best-fit parameters for the atmosphere fittings
on NS radius $R$ at different values of the NS mass,
$M/M_\odot =0.8, 1.0,\ldots 2.0$.
The thick
dashed lines delimit the ranges of the apparent temperature
and luminosity (upper and lower dashed curves correspond to
$M/M_\odot =2.0$ and 0.8, respectively).

{\bf Fig.~4}
Comparison of the surface temperatures inferred from the
BB (empty boxes) and hydrogen atmosphere (hatched boxes)
fittings of the spectra of \ns\ and the
Vela pulsar with NS cooling curves for three
EOS of the NS matter, two NS compositions, and different
models for additional frictional heating.
The cooling curves are taken from Van Riper et al.~(1995),
where more detailed explanations can be found.

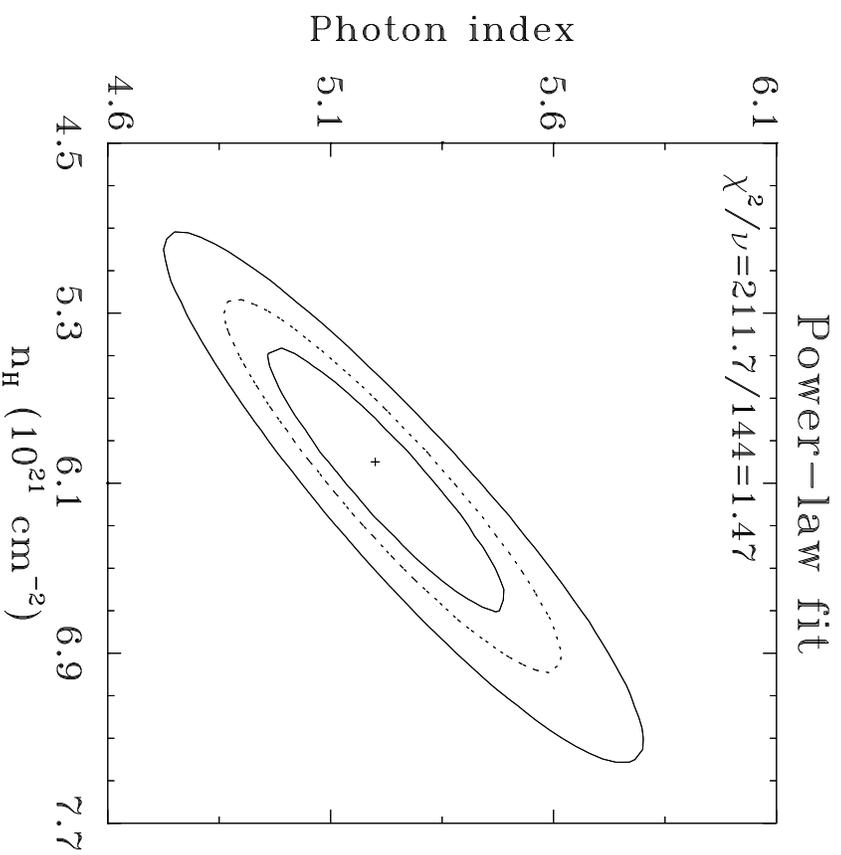
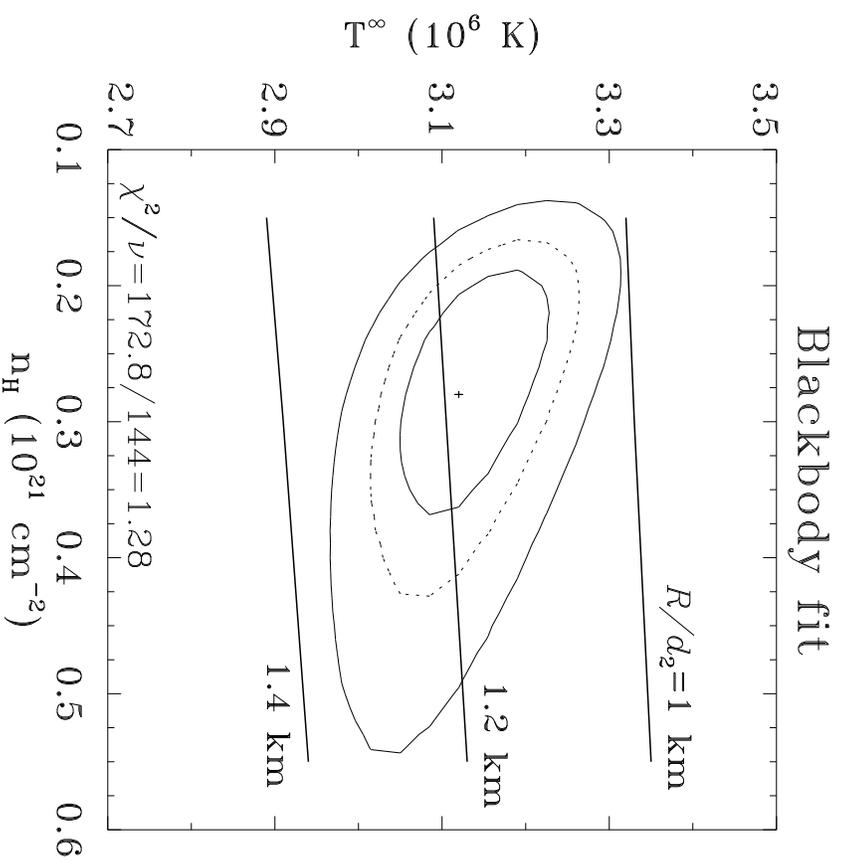
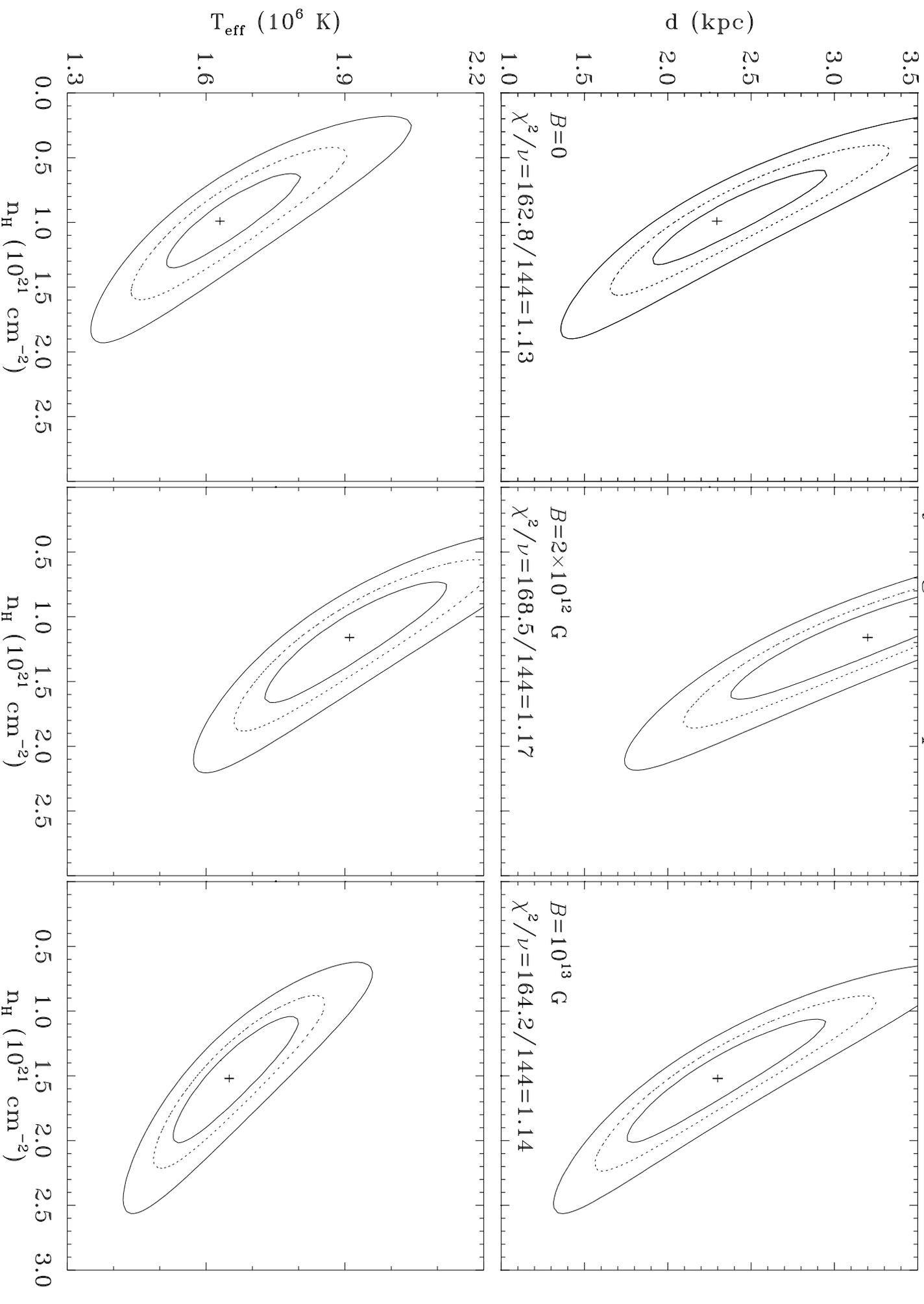
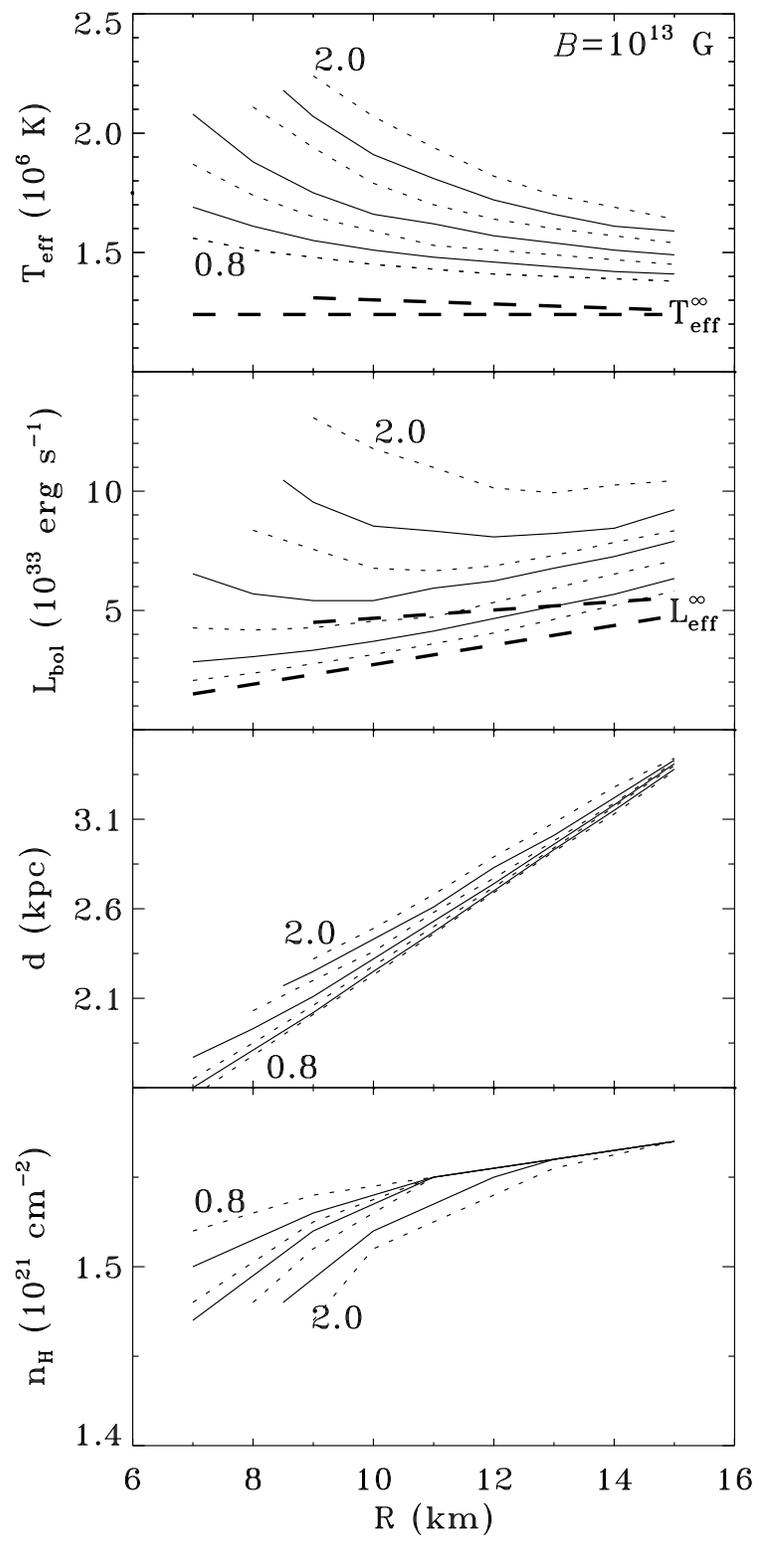
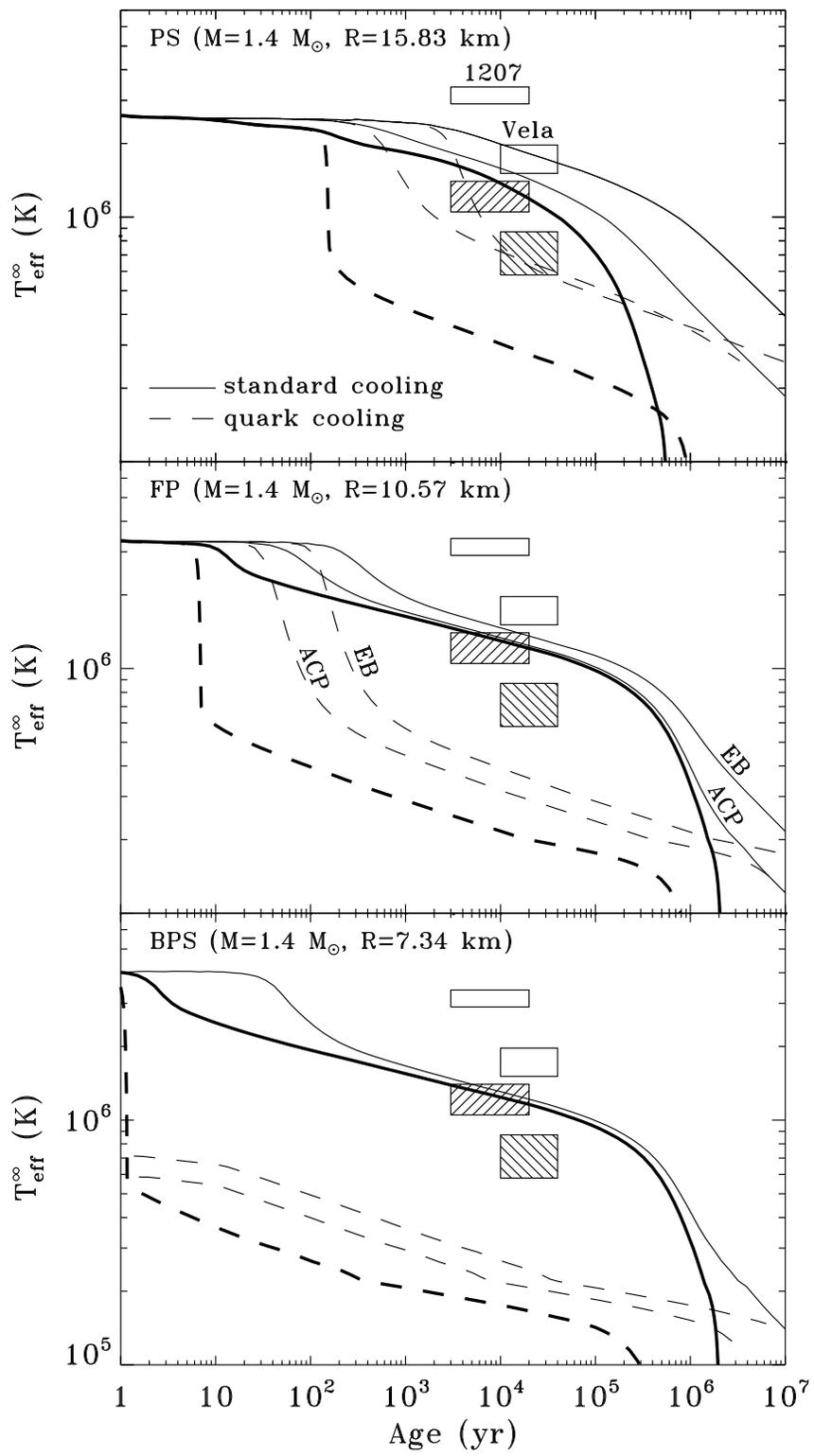
\end{document}